\documentclass[fleqn,usenatbib]{mnras}

\usepackage{newtxtext,newtxmath}

\usepackage[T1]{fontenc}
\usepackage{ae,aecompl}

\usepackage{graphicx}	
\usepackage{amsmath}	
\usepackage{amssymb}	

\def\kms{{\rm km~s$^{-1}$ }}

\title[Low X-Ray Luminosity Galaxy Clusters]
{Low X-Ray Luminosity Galaxy Clusters. IV. SDSS galaxy clusters at z $<$ 0.2\thanks {\dag: e-mail: anaomill@unc.edu.ar}
}
\author[O'Mill et al.]
 {
 \parbox[t]{\textwidth}{
Ana Laura O'Mill$^{1,2}$;
M. Victoria Alonso$^{1,2}$;
Carlos Valotto$^{1,2}$; and Jos\'e Luis Nilo Castell\'on$^{3,4}$}
\vspace*{6pt} \\
$^{1}$ Instituto de Astronom\'ia Te\'orica y Experimental, (IATE-CONICET), Laprida 922, C\'ordoba, Argentina.\\
$^{2}$ Observatorio Astron\'omico de C\'ordoba, Universidad Nacional de C\'ordoba, Laprida 854, C\'ordoba, Argentina.\\
$^{3}$ Departamento de F\'isica y Astronom\'ia, Facultad de Ciencias, Universidad de La Serena,  Av. Juan Cisternas 1200 Norte, La Serena, Chile.\\
$^{4}$ Direcci\'on de Investigaci\'on y Desarrollo, Universidad de La 
Serena, Av. Ra\'ul Bitr\'an Nachary 1305, La Serena, Chile.
}

 \date{\today}

\pubyear{2019}

\begin{document}
\label{firstpage}
\pagerange{\pageref{firstpage}--\pageref{lastpage}}
\maketitle

\begin{abstract}
This is the forth of a series of papers on low X-ray luminosity galaxy clusters.
The sample comprises 45 galaxy clusters with X-ray luminosities 
fainter than 0.7 10$^{44}$  erg s$^{-1}$ at redshifts lower than 0.2 in the regions of the Sloan Digital Sky Survey. The sample of spectroscopic members of the galaxy clusters was obtained with the criteria:  r$_p$ $\le$ 1 Mpc and 
$\Delta V \leq \sigma$ using our $\sigma$ estimates containing 21 galaxy clusters with more than 6 spectroscopic members.  
We have also defined a sample of photometric members with galaxies 
that satisfy r$_p \le $ 1 Mpc, and $\Delta V \leq $ 6000 \kms including 45 galaxy clusters with more than 6 cluster members.  
 We have divided the redshift range in three bins:  $z \leq  0.065$; 0.065 $<$ z $<$ 0.10; and z $\ge$ 0.10.  We have stacked the galaxy clusters using the spectroscopic sub-sample and we have computed the best RS linear fit within 1$\sigma$ dispersion.   With the photometric sub-sample we have added more data to the RS obtaining the photometric 1$\sigma$ dispersion relative to the spectroscopic RS fit.  
We have computed the luminosity function using the $1/V_{max}$ method  fitting it with a Schechter function. 
The  obtained  parameters for  these galaxy clusters with low X-ray luminosities are remarkably similar  to those for groups and poor galaxy clusters at these lower redshifts. 

\end{abstract}




\section{Introduction}

 The hierarchical model of structure formation predicts
that the progenitors of 
the most massive galaxy clusters are relatively small systems that are 
assembled together at high redshifts 
(\citealt{mcgee2009}; \citealt{delucia2010}).  Galaxy clusters of different masses provide physical insights into galaxy evolution. Less massive clusters show members with morphological properties similar to massive clusters at
the same redshift (\citealt{balogh2002}; \citealt{nilo14}).  

At lower redshifts, galaxy clusters contain a rich population of red 
early-type galaxies, lying on a tight relation, the cluster red sequence (RS) in the Colour-Magnitude Diagrams (CMD, \citealt{vis77}; \citealt{bower92a}; \citealt{bower92b}; \citealt{gladders1998}; \citealt{delucia2004}; \citealt{gil08}; \citealt{lerchster2011}).  
The changes in the slope and zero-point  might be an indication of the cluster evolution (\citealt{stott2009}). Also, star-forming, late-type galaxies populate the ''blue cloud'' in the CMDs. The presence of these two populations explains the observed bimodality in the colour distribution (\citealt{baldry2004}).  
The existence of the RS; the blue galaxy population; and the galaxy
interactions as a function of redshift and environment are important properties to characterize galaxy clusters. 
These properties can be used to test the formation models as the hierarchical merging model (\citealt{kauffmanncharlot1998}; \citealt{delucia2004}).

The galaxy luminosity function (LF) is a fundamental tool to understand the processes involved in the formation and evolution of galaxies and in particular, to assess how the environment may affect the baryonic component of the galaxies. The general analytic expression is the Schechter function (\citealt{schecter1976}). The observed LF strongly differs from that expected by assuming the Mass Function of dark matter halos  as predicted in the current Cold Dark Matter model (\citealt{moore1999}, \citealt{klypin1999}, \citealt{jenkins2001}).  In this scenario of structure formation, the slope of the mass function at the low-mass end is close to -1.8. However, the observed shallow slope of the field galaxy LF can be obtained in semi-analytical models that include star formation as feedback effect (\citealt{dekelsilk1986}, and \citealt{bower2006}) 
which  brings the mass function slope into agreement with that of the LF. 
 
One of the main controversy in LF studies has been its universal character or its dependence on environment.  The bright part has been determined fairly accurately in a wide variety of environments ranging from dense clusters to the field.  On the other hand, the faint-end dependence on environment has been subject to uncertainties and systematic issues, mainly associated with the 
methodologies, data sets and how membership is defined.  \cite{boue2008} minimized the background contamination using the 
 presence of the RS in Abell 496. 
 They found a global LF with a faint-end slope of -1.55, significantly 
shallower than previous estimates without colour cuts. The slope 
is shallower in the central region ($\alpha =-1.4$) and steeper in the envelope of the cluster ($\alpha$ = -1.8).
It has also been argued that different
processes associated with the cluster environment can 
contribute to enlarge or diminish the faint-end of the LF. The existence of an excess of dwarf galaxies in clusters would have important implications on galaxy formation and evolution models.
In the ''downsizing'' scenario, dwarf galaxies form or enter the clusters later than giant galaxies \citep{delucia2004} and this effect would depend on the cluster mass. Dwarf elliptical galaxies could also be the result of stripped discs. They suggest that tidal interactions could even cause the destruction of dwarf galaxies and consequently the reduction of faint galaxies. 

The LF of massive Abell clusters based on spectroscopic data shows a flat faint-end (e.g. \citealt{gaidos1997}; \citealt{paolillo2001}). 
Conversely, the LF of poor clusters has been studied mainly using photometric data. 
\cite{valotto1997} analyzing a large sample of galaxy clusters found that poor systems have flatter faint-end than rich ones.  \cite{lopezcruz1997} found that galaxy clusters without X-ray emission have a very steep faint-end while bright X-ray clusters have a flat behaviour.  This can be interpreted in terms of disruption of a large fraction of dwarf galaxies during the early stages of cluster evolution. 
Also analysing X-ray clusters, \cite{popesso2006} found that all the 
cluster LFs appear to have the same shape within the cluster physical sizes, with a marked upturn and steepening at the faint-end.  Using data from the Sloan Digital Sky Survey (\citealt{abazajian2009}, SDSS), 
\cite{defilippis2011} analysed the LF down to $M_r$ = -16 mag for a sample of galaxy clusters of the  Northern Sky Optical Cluster Survey at z $<$ 0.2.  They found a global LF with no  evidences for an up-turn at faint magnitudes. In this sense, it is crucial to define the galaxy cluster in terms of real members confirmed by spectroscopic measurements as pointed out by \cite{delapparent2003}.  The first attempt of using spectroscopy to improve the galaxy LF determinations  was done by \cite{christleinzabludoff2003}  in some 
nearby clusters reaching  brighter magnitudes, 
m$_R \approx$   17 mag (or M$_R$ = -17 mag). Finally, \citet{zandivarez2011} found that
galaxies in groups at low density regions have variations
on the Schechter parameters as a function of group mass: M$^*$ from -20.3 to -20.8 mag and $\alpha$ from -0.85 to -1.1. At high density regions, the groups have a
constant behaviour with values of M* $\approx$ -20.7 mag and $\alpha$ $\approx$ -1.1.

We intend to contribute to galaxy formation and evolution studying a sample of low X-ray luminosity galaxy clusters at intermediate redshifts.  The main goals of the project and the galaxy cluster sample is described in \citet{nilo16}. We have been interested mainly in the cluster galaxy populations and the assignment of the cluster membership at redshift range 0.16 $<$ z $<$ 0.7.   
The second paper of the series (\citealt{nilo14}) have presented the galaxy properties of the galaxy clusters obtained with Gemini data.  In the studied redshift range, the RS is clearly present in systems at lower redshifts while it is less important at higher redshifts.  It was also observed an increasing fraction of blue galaxies and a decreasing fraction of lenticulars, with a constant fraction of early-type galaxies with the cluster redshifts.
In the third paper (\citealt{gonzalez2015}), we have performed the weak lensing analysis of these galaxy clusters to estimate their masses. They correlate with the
observed M--L$_X$ relation and models.

We may take advantage of the homogeneous galaxy data using the available SDSS spectroscopic and photometric data. For a sample of galaxy clusters, the goal is to assign membership, to determine the RS and to obtain the LF.  In this forth paper, we are interested in the low X-ray luminosity galaxy clusters at lower redshifts in the SDSS region.  One of the key points here is to define
cluster members
and we use both spectroscopic and photometric redshift estimates.  
This paper is organized as follows: in section $\S$2 we define
the sample of galaxy clusters with low X-ray luminosities. In section $\S$3, we define the cluster member assignment using the spectroscopic and photometric SDSS data 
and we also divide the redshift range in three bins to stack the cluster data.
In section $\S$4, we obtain the cluster red sequence in the three redshift bins and we study the morphology of the cluster members.  In section $\S$5, we discuss the luminosity
function determinations stacking the galaxy clusters in the three redshift bins.  Finally in section $\S$6, we summarize the main results. 

Throughout this work, we adopt the cosmological model characterized by the  parameters: 
$\Omega_\Lambda$ = 0.7, $\Omega_m $ = 0.3 and $H_0$ = 75 h~km~s$^{-1~}$Mpc$^{-1}$.


\section{The Sample of low X-ray Luminosity Galaxy Clusters}

\citet{nilo16} have defined a sample of low X-ray luminosity galaxy clusters
based on the extended X-ray emission from the ROSAT Position Sensitive
Proportional Counters survey and the galaxy clusters selected by 
\citet{vikhlinin1998} and \citet{mull03}.  This sample includes 140 galaxy
clusters with X-ray luminosities in the [0.5--2.0] keV energy band (rest frame)
in the range of $10^{42}$ to $\sim 50 \times 10^{43}$ erg s$^{-1}$ and redshifts between 0.16 to 0.7.  The lower redshift limit was imposed by the field of view of the different instruments used to perform the photometry.  

In this work, we have selected \citet{mull03} low X-ray galaxy clusters within the same luminosity range
at redshifts lower than 0.2.  These galaxy clusters have available photometric and spectroscopic data of the 
Sloan Digital Sky Survey (\citealt{dr13}, hereafter SDSS-DR13\footnote {http://www.sdss.org/dr13/}).  
 The sample 
has a total of 45 galaxy clusters and through the work we refer them as the SDSS
galaxy clusters. [VMF98]178 is a galaxy cluster with strong
contamination by a foreground star with only a few photometric measurements
in the SDSS-DR13 and it is not considered in this analysis.

\begin{figure*}
\includegraphics[width=80mm]{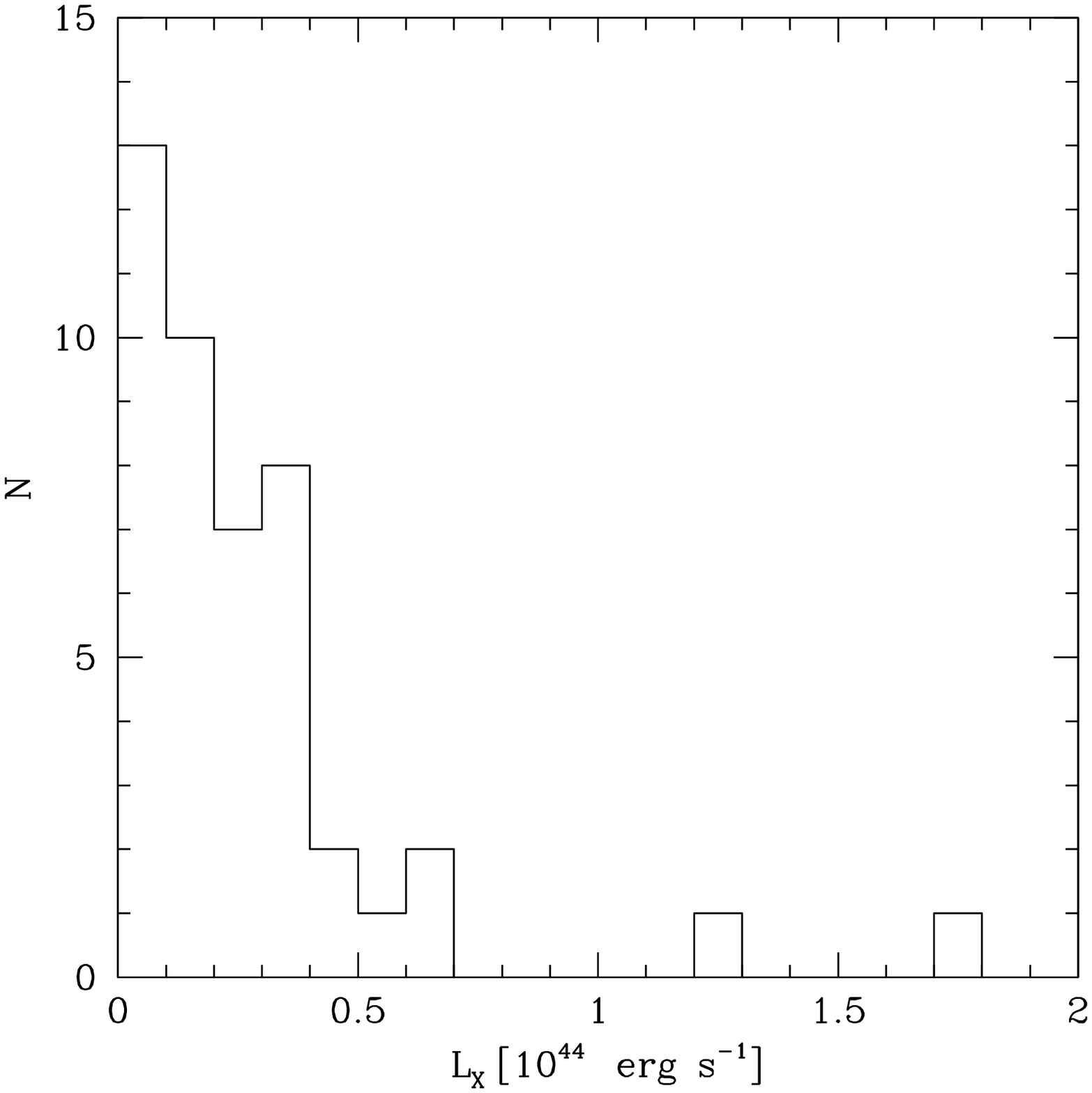}
\includegraphics[width=80mm]{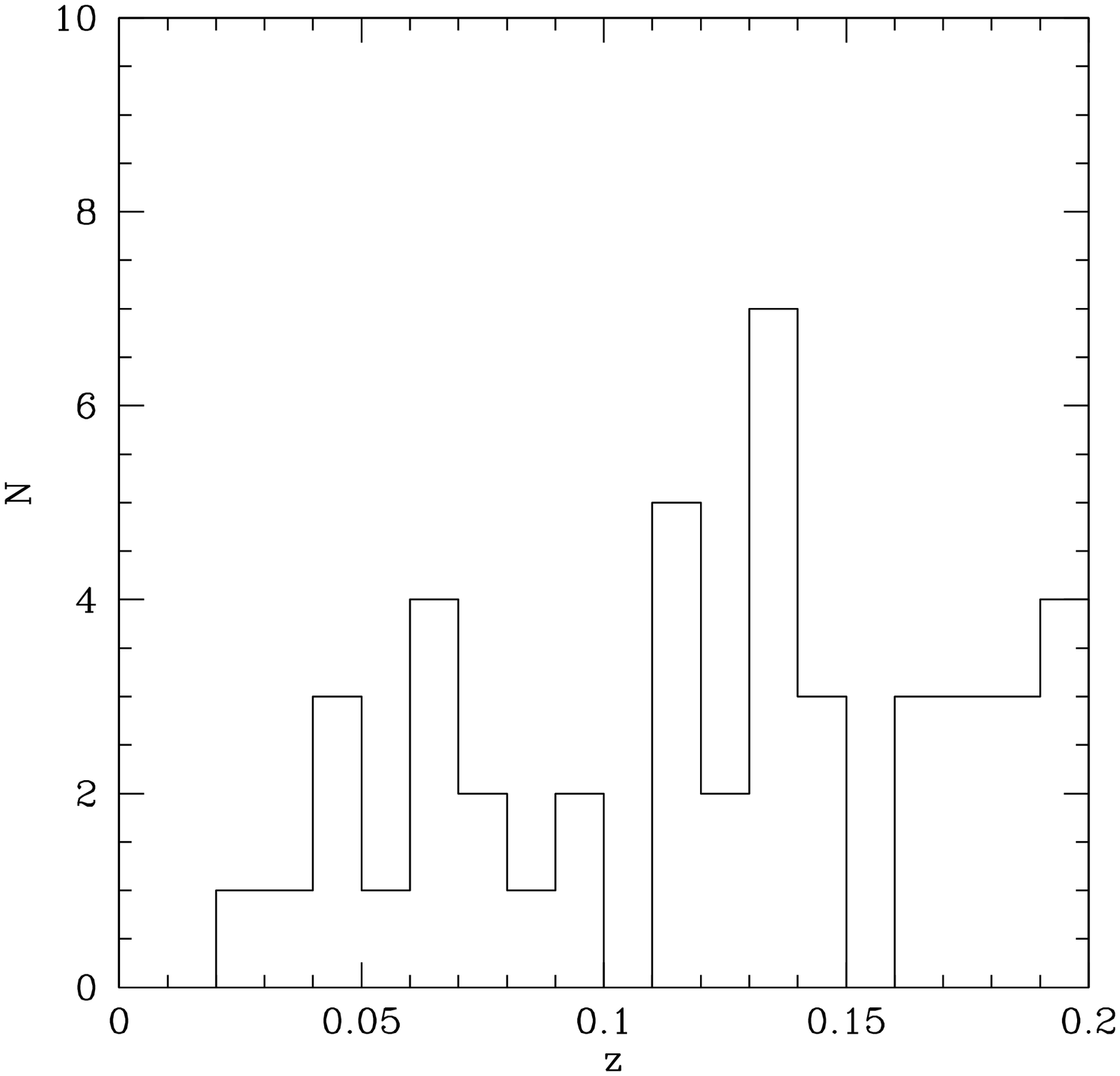}
\caption{X-ray luminosity and redshift distributions of the sample of 45 SDSS
  low X-ray galaxy clusters.}
\label{fig0}
\end{figure*}

Table~\ref{table1} shows the sample of SDSS galaxy clusters where columns (1)
and (2) are the \cite{vikhlinin1998} and ROSAT X-Ray survey 
identifications, respectively; columns (3) and (4), the J2000 equatorial coordinates
of the X-ray emission centroid; and columns (5) and (6), the X-ray luminosity in the [0.5--2.0] keV energy band  with estimates of the lower
bound of their uncertainties and the mean redshift (z$_M$) from \citet{mull03}, respectively. Figure~\ref{fig0} shows the X-ray luminosity and redshift distributions of the studied sample.
The galaxy clusters have X-ray luminosities
fainter than 0.7 10$^{44}$  erg s$^{-1}$ with only two brighter
clusters: [VMF98]112 and [VMF98]146 with X-ray luminosities of 1.24 and
1.80 10$^{44}$  erg s$^{-1}$, respectively.  In the figure, the number of galaxy clusters
grow with redshifts up to 0.15. 

\begin{figure*}
  \includegraphics[width=80mm]{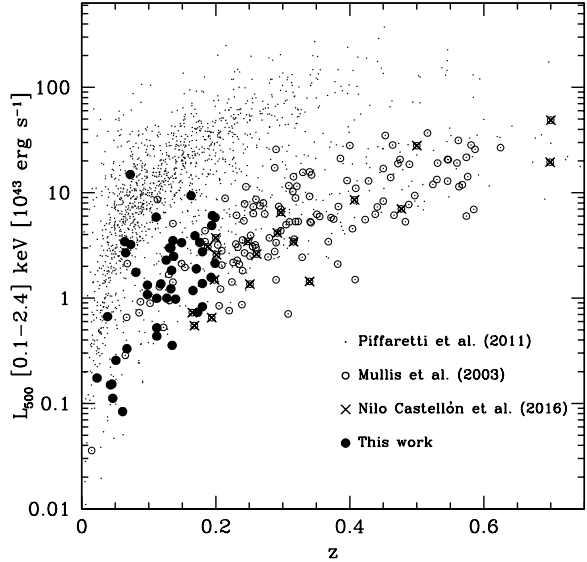}
  \caption{The X-ray luminosities and redshifts of the galaxy clusters.
    Small dots represent the \citet{piffaretti2011} compilation; open 
  circles, the \citet{mull03} sample;  crosses, \citet{nilo16} low X-ray galaxy clusters; and black circles, the SDSS galaxy clusters studied here.}
\label{fig1}
\end{figure*}

\cite{piffaretti2011} have presented the largest X-ray galaxy cluster compilation
based on the ROSAT All Sky Survey data.  Figure~\ref{fig1} shows the cluster
X-ray luminosities in the system of this compilation, and redshifts.
Small dots represent the galaxy cluster compilation and  open circles, the
\citet{mull03} galaxy clusters.  The galaxy clusters selected by \citet{nilo16} are highlighted with  crosses and the 45 SDSS galaxy clusters presented here with black circles.  

\section{The cluster member assignment}

In order to assign members to the SDSS galaxy clusters, we have selected galaxies with
spectroscopic
and photometric data from SDSS-DR13.  We have adopted the $ModelMag$ magnitudes corrected by extinction and then applied the offset (\citealt{doi2010}) and the k-correction
following
the empirical k-correction of \citet{omill11} at $z = 0.1$.  We have considered two galaxy samples:
the spectroscopic sample with magnitudes 14.5 $< r < $ 17.77 mag and the
photometric sample with $r < $ 21.5 mag.   We have only  considered galaxies with ($g - r$) $<$ 3 mag to minimize the inclusion of foreground stars (\citealt{collister2007}). 

\begin{figure*}
  \includegraphics[width=160mm]{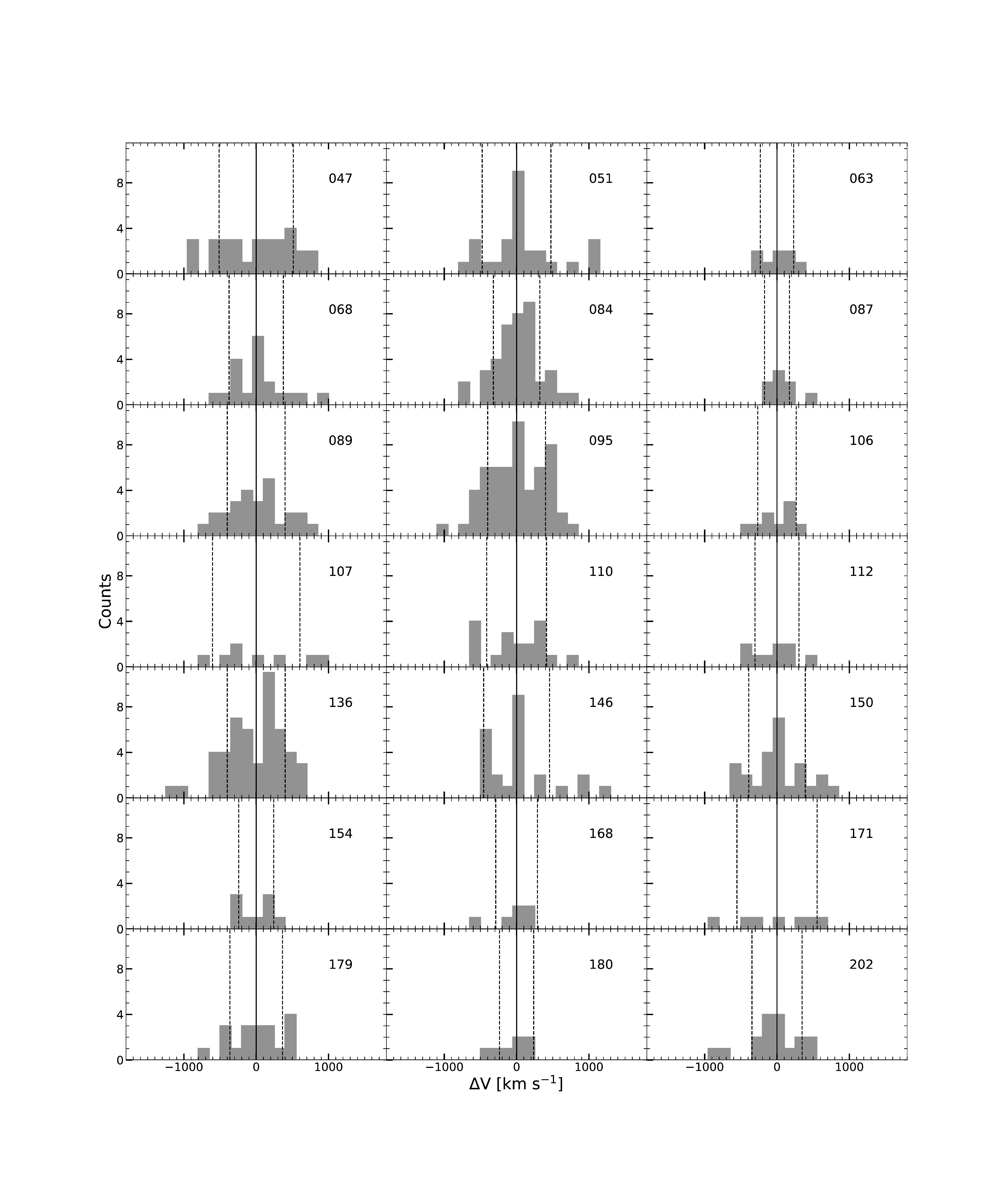}
\caption{$\Delta V $ distributions for the spectroscopic cluster sub-sample.  The [VMF98] cluster identification is included in each panel and our redshift estimates are shown with vertical solid lines and the velocity dispersions with dashed lines.}
\label{fig2}
\end{figure*}

For the spectroscopic sample, the galaxies were extracted from the $SpecObj$ table of the 
SDSS-DR13 using the projected radius, r$_p$, and
the radial velocity difference, $\Delta V$, relative to the cluster centre and cluster redshift, respectively. 
We have considered the centroid of the X-ray emission as the cluster
centre and we have  defined  r$_p$  $\le$ 1 Mpc, which is a typical cluster size.  In \citet{nilo16} we have
used a more restrictive condition (r$_p$  $\le$ 0.75 Mpc) choosing 
mainly the  central parts of the clusters due to the limitations of the field of view of the instruments.
To define our first guess of
cluster membership,  $\Delta V$ was calculated taking into account the \citet{mull03} cluster redshift and we have considered $\Delta V \leq $ 1000 \kms. We have dealt with  galaxy clusters with
more than 6 members defined with the r$_p$ and $\Delta V$ constraints to have a reliable characterization of the galaxy clusters.  With these galaxy clusters, we have also obtained new median cluster redshifts and bi-weight $\sigma$ estimates.
 The uncertainties were derived from a bootstrap resampling technique. Compared with \citet{mull03}  redshift estimates, the differences are smaller than 2 $\times 10^{-3}$, or about 600 \kms.  Using our redshift estimates as the new cluster redshifts, we have also obtained the cluster velocity dispersions based on the $''gapper''$ algorithm (\citealt{beers1990}). This is an efficient estimator less sensitive to outliers and  it reproduces accurately the true dispersion of the system.  Our two $\sigma$ measurements are equivalent with differences smaller than 10$^{-4}$. Table~\ref{table1} shows in columns (7) and (8), our new cluster redshifts  and  bi-weight velocity dispersion estimates  and their uncertainties, respectively.  
Using these velocity dispersions, we have defined the final spectroscopic cluster
members with $\Delta V$ relative to our new redshift estimates
with $\Delta V \leq \sigma $. This choice is in agreement with membership selection in galaxy clusters of \citet{biviano2013} and \citet{annunziatella2014}. Column (9) of Table~\ref{table1}
presents the number of spectroscopic cluster members assigned with our constraints.  The spectroscopic cluster sub-sample has 21 galaxy clusters with more than 6 spectroscopic members.  Figure~\ref{fig2} shows the $\Delta V$ distributions for
this spectroscopic sub-sample. The solid vertical lines represent our new cluster redshift estimates and dashed lines, $1\sigma$ of the velocity dispersion.

In order to improve the number of cluster members, we also selected galaxies without spectroscopic redshifts in the SDSS-DR13 using the $PHOTOZ$ table.  We have selected galaxies with  r$_p \le $ 1 Mpc, and $\Delta V \leq $ 10000 \kms due to higher photometric redshift  uncertainties.  We divided the total redshift range in this study in three bins to have the same number of galaxy clusters 
in each one:

\begin{itemize}
\item z$_1$: $z \leq  0.065$; 
\item z$_2$: 0.065 $<$ z $<$ 0.10; and
\item z$_3$: z $\ge$ 0.10.
\end{itemize}

\begin{figure*}
  \includegraphics[width=160mm]{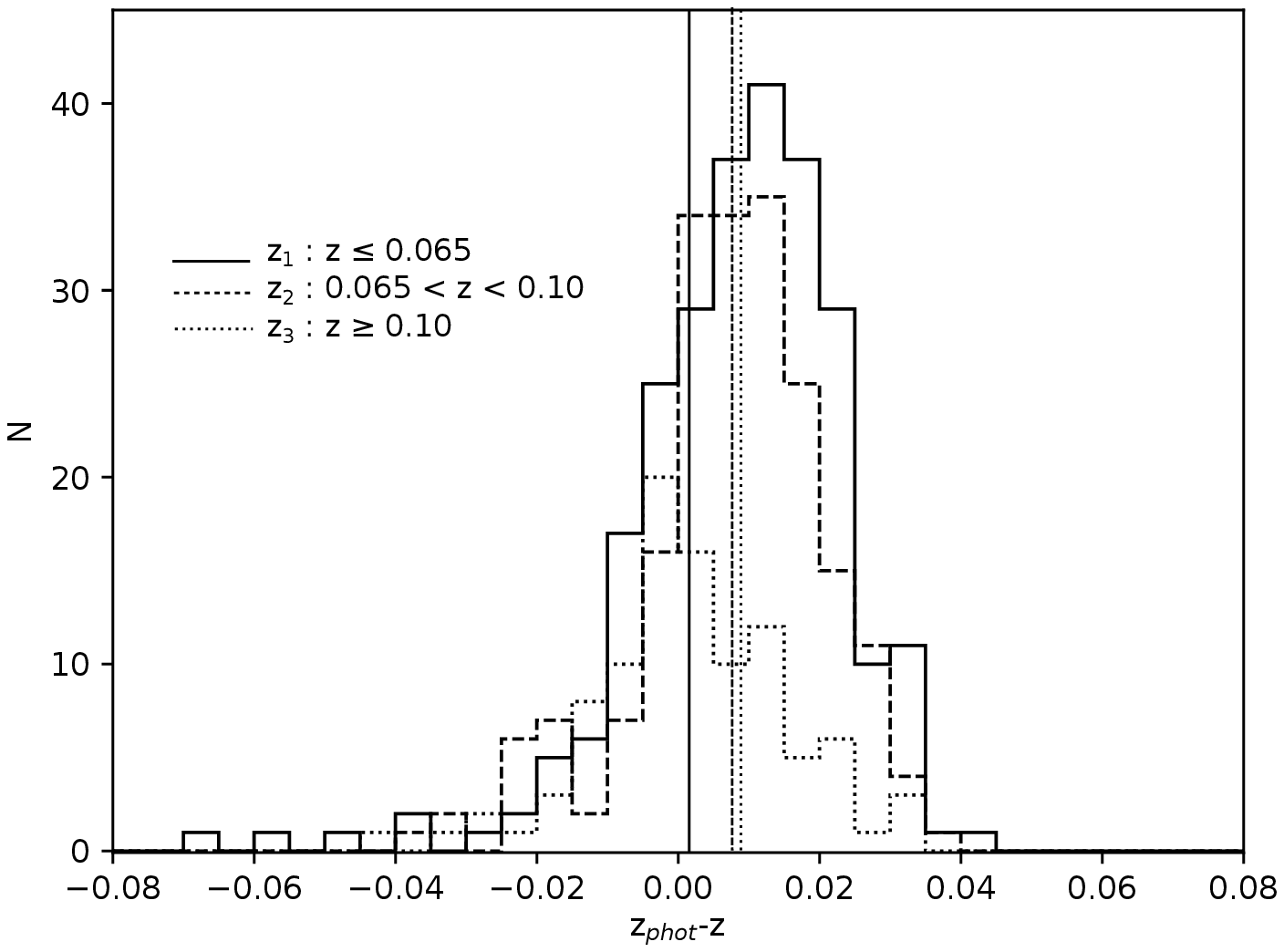}
\caption{The redshift difference distributions  between  
photometric  and  spectroscopic  redshift estimates in the three  bins.  Vertical lines represent the mean differences.  Both the distributions and mean differences are shown with different line types.}
\label{deltaz}
\end{figure*}

To investigate the uncertainties and possible biases in the photometric redshifts, we have made a comparison of galaxies with both spectroscopic and photometric redshift estimates for each bin.  
 Figure~\ref{deltaz} shows the distributions of redshift differences between photometric ($z_{phot}$) and spectroscopic ($z$) redshifts in the three defined bins.  Mean differences of $z_{phot} - z$ = 0.009 $\pm$ 0.014; 0.008 $\pm$ 0.013; and 0.002 $\pm$ 0.013 are also shown.  
The standard deviation of these values are around 0.01, which is equivalent to 6000 \kms.   Following \cite{costaduarte2016}, we have applied these differences to correct the photometric estimates to improve cluster membership.  This result is also in agreement with the cluster definition of \citet{nilo16}.
Using this procedure, the photometric sub-sample has 45 galaxy clusters with
more than 6 cluster members and the column (10) of the Table~\ref{table1} quotes the number of photometric members. 

Summarizing,  we have two samples of cluster members: the spectroscopic sample and the total sample that includes both spectroscopic and photometric members.  They are used
to study the cluster RS and the LF in the three redshift bins.

\section{The Cluster Red Sequence}

In the galaxy clusters, the early-type galaxies follow a well-defined relation in the Colour-Magnitude Diagrams extended by more than 4 magnitudes, the
cluster Red Sequence (RS,  \citealt{gladders1998}; \citealt{gladders2005}).  

\begin{figure*}
\includegraphics[width=160mm]{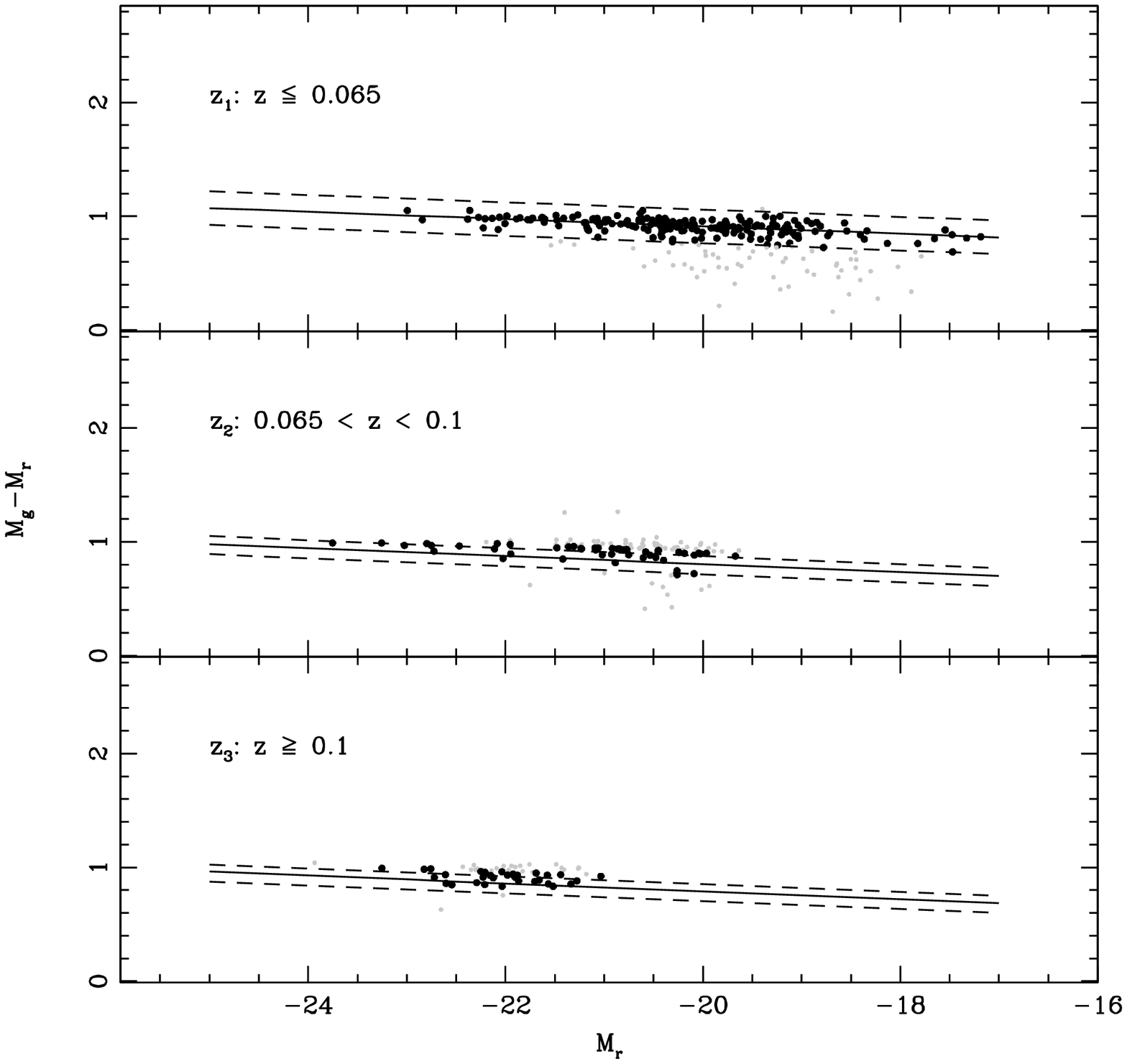}
\caption{Colour-Magnitude Diagrams for spectroscopic members of the galaxy
  clusters in the three redshift bins.  Grey points correspond to all spectroscopic members.  Solid lines show the best spectroscopic RS fits and dashed lines, the 1 $\sigma$ spectroscopic dispersions. Black points represent the galaxies within 1 $\sigma$ dispersion of the spectroscopic RS fits.}
\label{fig4}
\end{figure*}

In this work, we have considered absolute magnitudes of the galaxy cluster members and we have stacked all of them in the three redshift bins. 
Figure~\ref{fig4} shows the Colour-Magnitude Diagrams 
(M$_g$ - M$_r$ vs M$_r$) for the spectroscopic sub-sample of galaxy clusters in 
these bins.   
 We have also computed the best spectroscopic RS linear fit in z$_1$ and z$_2$.  In the case of the z$_3$ bin, we have fewer points only covering an interval 
 of about -23 to -21 in M$_r$ absolute magnitudes.  For this reason, we have considered the RS fit obtained for  z $<$ 0.1. Table~\ref{table2} shows the three redshift bins in column (1); the best linear fit spectroscopic RS
slopes and zero-points in columns
(2) and (3), respectively; and  the  1 $\sigma$ dispersion of the spectroscopic fits in column (4).  In each panel of the figures, grey points are all spectroscopic members, the solid
line represents the best spectroscopic RS linear fit quoted in the table
and the two dash lines are
the 1 $\sigma$ spectroscopic dispersions.  Black points represent the cluster
members that follow the spectroscopic RS fit within the dispersion.   For z$_1$ and z$_2$, the RS slopes are similar.   

\begin{figure*}
\includegraphics[width=160mm]{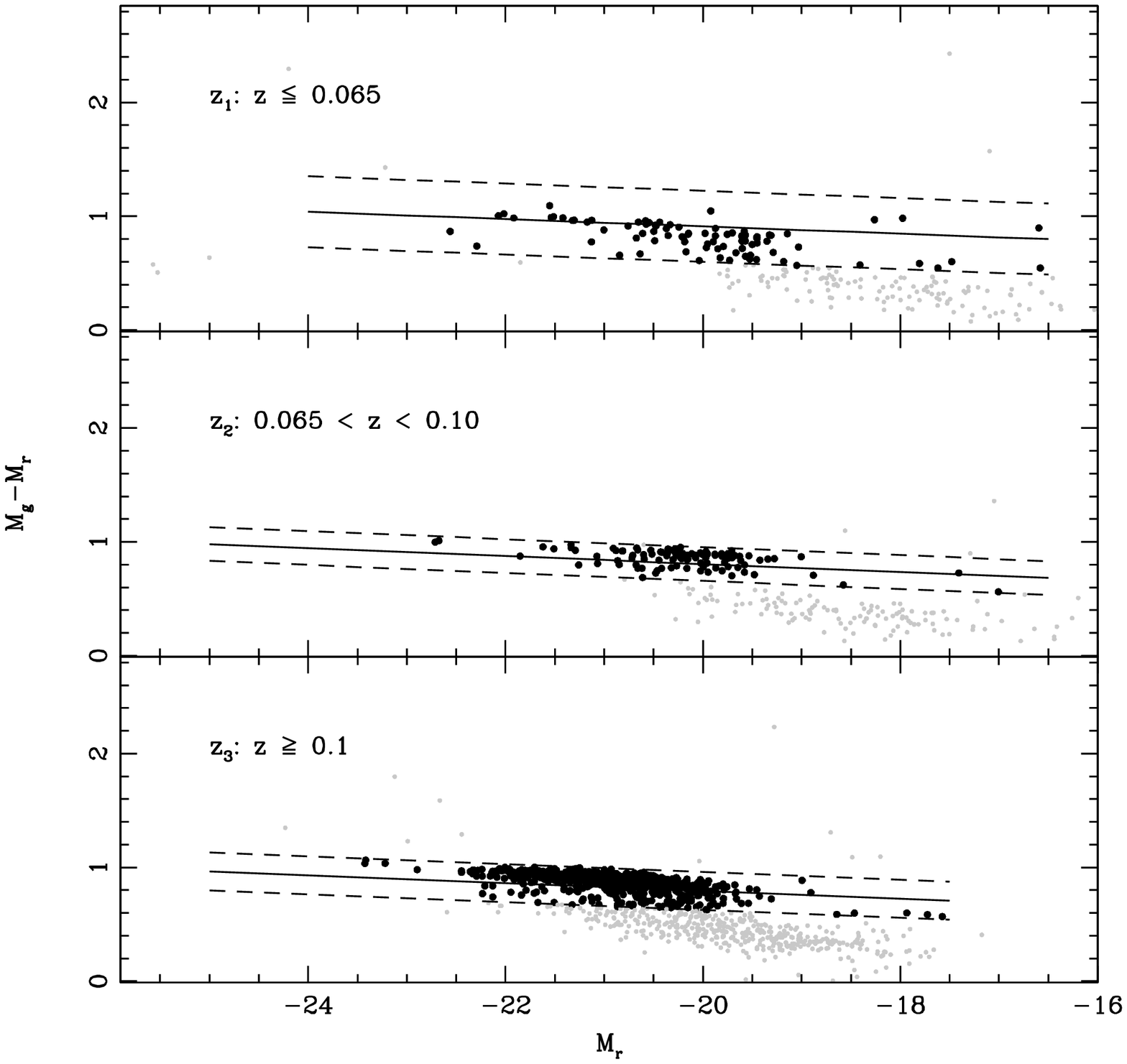}
\caption{Colour-Magnitude Diagrams for photometric members of the galaxy
  clusters in the three redshift bins. Grey points correspond to all the photometric
  galaxies.  Solid line shows the best spectroscopic RS fits and dashed lines are the 1 $\sigma$ photometric dispersion. Black points are the galaxies within 1 $\sigma$ photometric dispersion
  of the spectroscopic RS fits.}
\label{fig5}
\end{figure*}

We have also stacked the galaxy
clusters using the photometric members.  We have considered the RS fit
obtained with the spectroscopic members because of the higher uncertainties in the 
photometric estimates.  We have taken into account the spread of the photometric members relative to the spectroscopic RS fit and  
we have obtained the 1 $\sigma$ photometric dispersion, which is shown in the column (5) of the
Table~\ref{table2}.  Figure~\ref{fig5} shows the Colour-Magnitude Diagrams
(M$_g$ - M$_r$ vs M$_r$) for the galaxy clusters with  
photometric members in the three redshift bins. Grey points represent these photometric members; solid lines, the spectroscopic RS fits (table~\ref{table2}) 
and the two dash lines, the 1 $\sigma$ photometric dispersions.  Black points are the members that fit the spectroscopic RS within 1 $\sigma$ photometric dispersion. From the two figures we can see the passive galaxies that populate the inner parts of the galaxy clusters within 1$\sigma$ of the RS and  below it, the forming galaxy population. 

\begin{figure*}
\includegraphics[width=160mm]{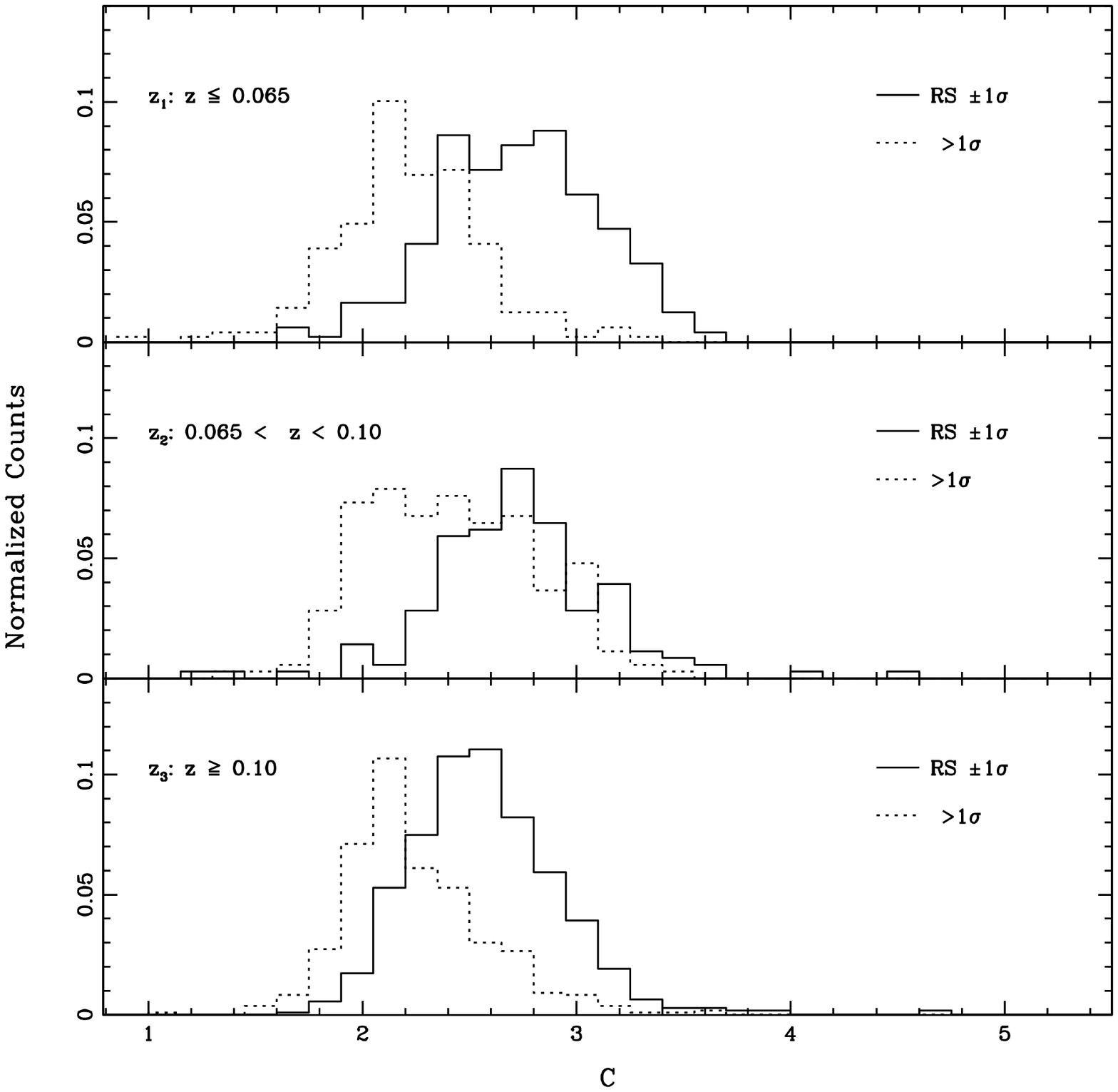}
\caption{Normalized $C$ distributions for the total sample of members in the 
three redshift bins.  Solid and dash histograms represent the distributions for members within 
$1\sigma$ of the spectroscopic RS fit and  below  it, respectively.  
}
\label{fig7}
\end{figure*}

To analyse the photometric properties of these two populations, we have also used $C$, the galaxy  concentration index. In the SDSS, this parameter is defined as the ratio of Petrossian radii that encloses
90 and 50 percent of the
galaxy light (\citealt{stout}).   
This quantity does not 
depend on distances and it is a suitable indicator of galaxy morphology (\citealt{strateva2001,kauff3a}; \citealt{kauff3b}).  Early-type 
galaxies have $C > 2.55$ and late-type galaxies, $C < 2.55$ (\citealt{strateva2001}). We have obtained $C$ values in the $r-band$ for the total sample of members. Figure~\ref{fig7} shows the normalized $C$ distributions of these members in the
three redshift bins.   Solid histograms represent the
distributions for galaxies within $1\sigma$ of the best spectroscopic RS fit 
and dash histograms, the distributions of those galaxies  below the RS. 
The two $C$ distributions are different and we fit Gaussian functions  to these  two populations:  early; and late types. Table~\ref{table3} shows 
the mean $C$ values for the two populations in the different redshift bins. The mean $C$s are in agreement with the above definitions of galaxy types:   about 2.7 for galaxies within $1\sigma$ the spectroscopic RS that correspond to early-type galaxies; and about 2.2 to late-types. At lower redshifts (z$_1$), the $C$ distribution of early-types is clear bimodal suggesting
the presence of an intermediate galaxy population.  Fitting two Gaussians we obtained $C$ = 2.50 $\pm$ 0.06 for this population. The results of the RSs and $C$ distributions suggest that  in this work, we minimize the number of outliers in the sample of cluster members.  
 
\section{The galaxy luminosity function}

The Schechter luminosity function (\citealt{schecter1976}) provides a parametric description of the space density of galaxies as a function of their luminosity.  The form of this function is:

\begin{equation*}
N(M)=0.4~ln(10)~\Phi^*(10^{0.4(M^*-M)})^{(\alpha+1)}exp(-10^{0.4(M^*-M)})
\end{equation*}

with M$^*$, the characteristic absolute magnitude that corresponds to the ''knee'' of the function; the slope $\alpha$ of the power law that dominates the faint-end; and $\phi^*$, the characteristic density. 
By using the cluster members, the computation of the 
cluster LF is straightforward without the need to obtain background galaxy determinations only requiring a correction for incompleteness.

The LF was obtained stacking the galaxy clusters in the three different redshift bins.  
We  have computed the LF using the $1/V_{max}$ method considering the incompleteness with a $V/V_{max}$ test (\citealt{vmax}).
This method takes into account the volume of the survey enclosed by the galaxy redshift and the difference between the maximum and minimum volumes within which it can be observed. 
The value $\langle V/V_{max} \rangle$ varies with apparent magnitude.  
Considering a sample of sources uniformly distributed, we expect that  half will be found in the inner half of the volume $V_{max}$ and the rest in the outer half. On average, we expect that $\langle V/V_{max} \rangle \sim 0.5$,  and in this way, we can define the limiting magnitude where the catalogue is complete. In our case, we reached apparent magnitudes of $\sim 19.8$ mag in the $r-band$, which is a good compromise between the
limiting magnitudes of both the spectroscopic and photometric samples.  The LF errors were computed using the bootstrap re-sampling technique.  

\begin{figure*}
\includegraphics[width=160mm]{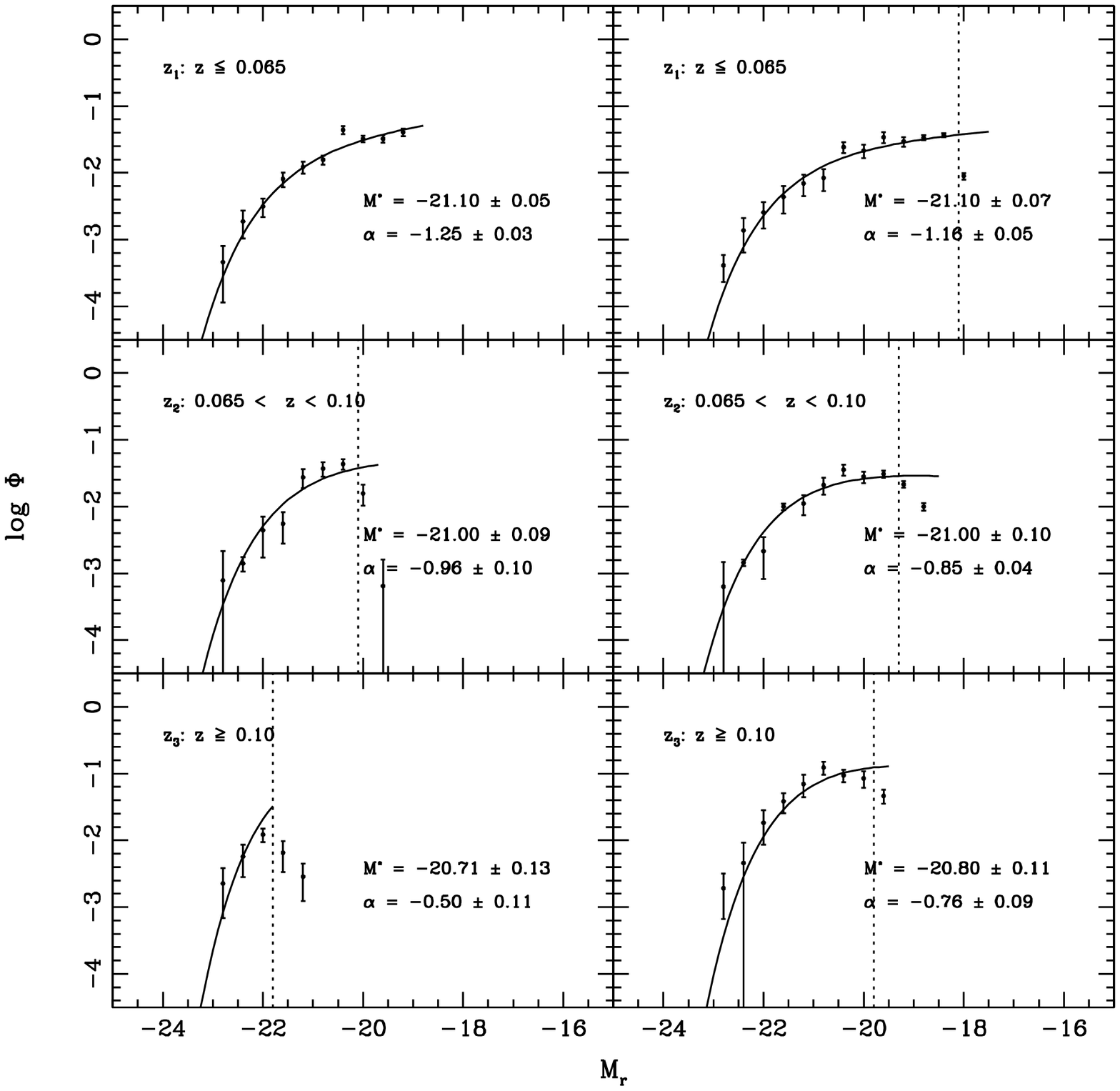}
\caption{Stacked Luminosity Functions in the three redsfhit bins.  Left and right panels show the LFs obtained with only spectroscopic members and with the total sample of members, respectively.  Vertical lines represent the limiting magnitudes beyond which the points were not considered in the Schechter fit and solid lines represent the best Schechter fit. }
\label{fl}
\end{figure*}

Figure~\ref{fl} shows the stacked LF and errors for the three redshift bins  using only spectroscopic members (left panels) and the total sample of members (right panels).  Vertical lines represent the limiting magnitudes of the points considered in the Schechter fit.   Solid 
lines represent the Schechter LF fit and the $M^*$ and $\alpha$ parameters of the fit are also shown in each panel.  At lower redshifts, we can see that the LFs reach fainter magnitudes compared with the other two redshift bins.  In general, the photometric members have a strong
contribution at the LF faint-end in the three redshift bins.
At higher redshifts, the number of spectroscopic members is smaller and they only contribute to the bright part.  Including the photometric members allow us to  extend the LF at the faint-end. 
 Table~\ref{table4} presents the three Schechter LF parameters in the studied redshift bins using spectroscopic members in columns (2) to (4); 
and with the full sample of members in columns (5) to (7).

Our $M^*$ and $\alpha$ results are consistent with other authors using the SDSS data  in the $r-band$.  For instance, \cite{zandivarez2011} obtained Schechter parameters for LFs in groups with mass ranging 12 $\le$ $log({\cal M}/M_{\odot}$ h$^{-1}$) $\le$ 15 up to redshifts of 0.2.  \citet{goto2002} converted parameters to this passband finding that $M^*$ ranges from -20.00 to -22.55 mag and $\alpha$ from -0.69 to -1.4 for different galaxy clusters (\citealt{valotto1997}, \citealt{garilli1999} and \citealt{paolillo2000}).  Their estimates have higher uncertainties because they obtained the LF with only photometric data and they  applied the statistical method for background subtraction. Our results of galaxy clusters with low X-ray luminosities are consistent with all these previous studies of groups and loose or poor galaxy clusters.  The properties of these
systems might be different but the parameters of the LFs are remarkably similar. This result suggests that the LF is a generic feature of the galaxy systems.  In the redshift range of our study, the results are also in agreement with the study of \citet{sarron2018} with SDSS galaxy clusters at z $<$ 0.7.  They found similar $M^*$ and $\alpha$ parameters in their range of redshifts and
masses.

\section{Summary}

This work is the forth in a series of papers
aimed at understanding the processes involved in the formation and
evolution of low X-ray luminosity galaxy clusters at intermediate
redshifts.  

The sample of spectroscopic members was defined with galaxies from the SDSS-DR13 that follow the criteria: r$_p$  $\le$ 1 Mpc and 
$\Delta V \leq \sigma$ using our $\sigma$ measurements.
We have obtained a sub-sample of 21 galaxy clusters with more than 6 spectroscopic members.  
We have also defined a sample of photometric members with galaxies 
with r$_p \le $ 1 Mpc, and $\Delta V \leq $ 6000 \kms.  This photometric sub-sample has 45 galaxy clusters with
more than 6 cluster members.  

We have divided the redshift range in three bins:  $z \leq  0.065$; 0.065 $<$ z $<$ 0.10; and z $\ge$ 0.10.
We have stacked the galaxy clusters in the three redshift bins using the spectroscopic sub-sample and we have computed the best RS linear fit within 1$\sigma$ dispersion.  With the photometric sub-sample we
have added more data to the RS obtaining the photometric 1$\sigma$ dispersion relative to the spectroscopic RS fit.   The $C$ distributions obtained with the members within 1$\sigma$ RS fit and  below it  have mean values of about 2.7 (early-type galaxies) and about 2.2 (late-type galaxies). These results are in agreement with 
the idea that early-type galaxies are the responsible of the tight RS relation. 

We have also stacked the galaxy clusters in the three different redshift bins and computed the luminosity function using the $1/V_{max}$ method (\citealt{vmax}).   
The Schechter parameters obtained for galaxy clusters with low X-ray luminosities are remarkably similar to other authors, in particular to groups and poor galaxy clusters at these lower redshifts.  

We plan to extend this study to our complete sample of galaxy clusters at low X-ray luminosities in the redshift range 0.2 $<z<$ 0.7.

\vskip 0.5cm

\section*{Acknowledgements}

We would like to thank the referee, Florence Durret, for providing us with helpful
comments that improved this study. The work was mainly supported by the Ministerio de Ciencia y Tecnolog\'ia (GRFT 2017, No. 39, C\'ordoba) and partially by the 
Consejo de Investigaciones Cient\'ificas y T\'ecnicas (CONICET) and Secretar\'ia
de Ciencia y T\'ecnica (Secyt) of the Universidad Nacional de C\'ordoba.  JLNC is also
grateful for financial support received from the 
Programa de Incentivo a la 
Investigaci\'on Acad\'emica de la Direcci\'on de Investigaci\'on de la 
Universidad de La Serena (PIA-DIULS), Programa DIULS de Iniciaci\'on 
Cient\'ifica No. PI15142. JLNC also acknowledges the 
financial support from the 
GRANT PROGRAM No. FA9550-18-1-0018 of the Southern Office of Aerospace 
Research and Development (SOARD), a branch of the Air Force Office of the
Scientific Research's International Office of the United States (AFOSR/IO).

SDSS-XIII is managed by the Astrophysical Research Consortium for the Participating Institutions of the SDSS-III Collaboration including the University of Arizona, the Brazilian Participation Group, Brookhaven National Laboratory, University of Cambridge, University of Florida, the French Participation Group, the German Participation Group, the Instituto de Astrof\'isica de Canarias, the Michigan State/Notre Dame/JINA Participation Group, Johns Hopkins University, Lawrence Berkeley National Laboratory, Max Planck Institute for Astrophysics, New Mexico State University, New York University, Ohio State University, Pennsylvania State University, University of Portsmouth, Princeton University, the Spanish Participation Group, University of Tokyo, University of Utah, Vanderbilt University, University of Virginia, University of Washington, and Yale University.



\begin{table*}
	\centering
	\caption{The sample of low X-ray luminosity galaxy clusters. Column (1) shows
the \citet{vikhlinin1998} identification; column (2), the ROSAT X-Ray survey 
identification; columns (3) and (4), the J2000 equatorial coordinates
of the X-ray emission centroid; column (5), the X-ray luminosity in the [0.5--2.0] keV energy band and uncertainties; column (6),  the mean redshift  from \citet{mull03}; column (7),
our new cluster redshifts; column (8),  the bi-weight velocity dispersion estimates and uncertainties; column (9), the number of spectroscopic cluster members assigned with our constraints; and column (10), the number of photometric members.}
	{\small
 	\begin{tabular}{lllrcccccc}
		\hline
		\hline

[VMF98] & ROSAT X-Ray & Right Ascension & Declination &  L$_{X}$ &   z$_M$     & z & $\sigma$ & \# spec    & \# phot     \\
 Id.    & Id.   &   (J2000)     &  (J2000)      & (10$^{44}$  erg s$^{-1}$) &          & & & members  & members        \\  
		\hline                                                                     & &           &	&	  	  & &     &	&	  &	  \\
002    &	RX J0041.1-2339	 &    00 41 10.3  &  -23 39 33    &   0.06  $\pm$ 0.01  &  0.112 & &   &  -  & 12 \\

017    &	RX J0139.9+1810	 &    01 39 54.3  &  +18 10 00    &   0.38  $\pm$ 0.05 &  0.177 & &   &  -  & 22  \\
019    & 	RX J0144.4+0212	 &    01 44 29.1  &  +02 12 37    &   0.13  $\pm$ 0.03 &  0.166 & &   &  -  & 23  \\
031    &	RX J0259.5+0013	 &    02 59 33.9  &  +00 13 47    &   0.54  $\pm$ 0.04 &  0.194 & &   &  -  & 25  \\
047    &	RX J0810.3+4216	 &    08 10 23.9  &  +42 16 24    &   0.42  $\pm$ 0.05 &  0.064 & 0.064 & 513 $\pm$ 91  & 30  & 16  \\
051    & 	RX J0820.4+5645	 &    08 20 26.4  &  +56 45 22    &   0.02  $\pm$ 0.01 &  0.043 & 0.044 & 475 $\pm$ 71  & 27  & 21  \\
063    &	RX J0852.5+1618	 &    08 52 33.6  &  +16 18 08    &   0.16  $\pm$ 0.03 &  0.098 & 0.099 & 230 $\pm$ 84  &  8  & 15  \\
068    &	RX J0907.3+1639	 &    09 07 20.4  &  +16 39 09    &   0.34  $\pm$ 0.04 &  0.073 & 0.073 & 375 $\pm$ 68  & 19  & 32  \\
074    &	RX J0943.7+1644	 &    09 43 44.7  &  +16 44 20    &   0.31  $\pm$ 0.06 &  0.180 & &   &  -  & 32  \\
084    &	RX J1010.2+5430	 &    10 10 14.7  &  +54 30 18    &   0.02  $\pm$ 0.01 &  0.047 & 0.046 & 320 $\pm$ 41  & 40  & 21  \\
087    & 	RX J1013.6+4933	 &    10 13 38.4  &  +49 33 07    &   0.36  $\pm$ 0.08 &  0.133 & 0.133 & 173 $\pm$ 49  &  8  & 12  \\
089    &	RX J1033.8+5703	 &    10 33 51.9  &  +57 03 10    &   0.01  $\pm$ 0.01 &  0.046 & 0.046 & 400 $\pm$ 74  & 26  & 18  \\
094    & 	RX J1056.2+4933	 &    10 56 12.6  &  +49 33 11    &   0.23  $\pm$ 0.03 &  0.199 & &   &  -  & 13  \\
095    & 	RX J1058.2+0136	 &    10 58 13.0  &  +01 36 57    &   0.09  $\pm$ 0.01 &  0.040 & 0.039 & 400 $\pm$ 52  & 55  & 12  \\
099    &	RX J1119.7+2126	 &    11 19 43.5  &  +21 26 44    &   0.01  $\pm$ 0.01 &  0.061 & &   &  -  &  6  \\
103    & 	RX J1124.6+4155	 &    11 24 36.9  &  +41 55 59    &   0.67  $\pm$ 0.15 &  0.195 & &   &  -  & 20  \\
104    &	RX J1135.9+2131	 &    11 35 54.5  &  +21 31 05    &   0.14  $\pm$ 0.03 &  0.133 & &   &  -  & 27  \\
105    &  	RX J1138.7+0315	 &    11 38 43.9  &  +03 15 38    &   0.11  $\pm$ 0.03 &  0.127 & &   &  -  & 17  \\
106    & 	RX J1142.0+2144	 &    11 42 04.6  &  +21 44 57    &   0.35  $\pm$ 0.13 &  0.131 & 0.129 & 266 $\pm$ 92 &  9  & 38  \\
107    &	RX J1146.4+2854	 &    11 46 26.9  &  +28 54 15    &   0.38  $\pm$ 0.06 &  0.149 & 0.150 & 604 $\pm 197$  &  8  & 22  \\
109    & 	RX J1158.1+5521	 &    11 58 11.7  &  +55 21 45    &   0.04  $\pm$ 0.01 &  0.135 & &   &  -  & 10  \\
110    &	RX J1159.8+5531	 &    11 59 51.2  &  +55 31 56    &   0.21  $\pm$ 0.02 &  0.081 & 0.080 & 414 $\pm$ 94  & 18  & 15  \\
112    & 	RX J1204.0+2807	 &    12 04 04.0  &  +28 07 08    &   1.24  $\pm$ 0.14 &  0.167 & 0.166 & 304 $\pm$ 90  &  9  & 42  \\
126    &	RX J1254.6+2545	 &    12 54 38.3  &  +25 45 13    &   0.17  $\pm$ 0.03 &  0.193 & &   &  -  & 20  \\
134    & 	RX J1325.2+6550	 &    13 25 14.9  &  +65 50 29    &   0.15  $\pm$ 0.05 &  0.180 & &   &  -  & 24  \\
136    &	RX J1329.4+1143	 &    13 29 27.3  &  +11 43 31    &   0.02  $\pm$ 0.01 &  0.024 & 0.023 & 400 $\pm$ 53  & 50  & 24  \\
140    & 	RX J1336.7+3837	 &    13 36 42.1  &  +38 37 32    &   0.09  $\pm$ 0.02 &  0.180 & &   &  -  & 23  \\
144    &	RX J1340.5+4017	 &    13 40 33.5  &  +40 17 47    &   0.21  $\pm$ 0.03 &  0.171 & &   &  -  & 8  \\
145    &	RX J1340.8+3958	 &    13 40 53.7  &  +39 58 11    &   0.44  $\pm$ 0.04 &  0.169 & &   &  -  & 15  \\
146    & 	RX J1341.8+2622	 &    13 41 51.7  &  +26 22 54    &   1.80  $\pm$ 0.14 &  0.072 & 0.074 & 456 $\pm$ 72  & 24  & 32  \\
149    &	RX J1343.4+4053	 &    13 43 25.0  &  +40 53 14    &   0.11  $\pm$ 0.02 &  0.140 & &   &  -  & 6  \\
150    & 	RX J1343.4+5547	 &    13 43 29.0  &  +55 47 17    &   0.04  $\pm$ 0.01 &  0.069 & 0.068 & 390 $\pm$ 72  & 25  & 13  \\
154    &  	RX J1406.9+2834	 &    14 06 54.9  &  +28 34 17    &   0.16  $\pm$ 0.02 &  0.118 & 0.117 & 242 $\pm$ 84  &  9  & 22  \\
164    & 	RX J1438.9+6423	 &    14 38 55.5  &  +64 23 44    &   0.25  $\pm$ 0.03 &  0.146 & &   &  -  & 27  \\
168    & 	RX J1515.5+4346	 &    15 15 32.5  &  +43 46 39    &   0.29  $\pm$ 0.03 &  0.137 & 0.135 & 288 $\pm$ 74  &  6  & 21  \\
171    & 	RX J1537.7+1200	 &    15 37 44.3  &  +12 00 26    &   0.21  $\pm$ 0.06 &  0.134 & 0.133 & 555 $\pm$ 180 &  7  & 23  \\
173    & 	RX J1544.0+5346	 &    15 44 05.0  &  +53 46 27    &   0.05  $\pm$ 0.01 &  0.112 & &   &  -  & 5  \\

175    & 	RX J1552.2+2013	 &    15 52 12.3  &  +20 13 45    &   0.40  $\pm$ 0.05 &  0.136 & &   &  -  & 33  \\
177    & 	RX J1620.3+1723	 &    16 20 22.0  &  +17 23 05    &   0.12  $\pm$ 0.02 &  0.112 & &   &  -  & 20  \\

179    &        RX J1630.2+2434	 &    16 30 15.2  &  +24 34 59  &   0.33 $\pm$ 0.05 &  0.066 & 0.065 & 363 $\pm$ 82 & 19  & 21  \\
180    & 	RX J1631.0+2122	 &    16 31 04.6  &  +21 22 02    &   0.12 $\pm$ 0.03 &  0.098 & 0.096 & 237 $\pm$ 72  &  7  & 29  \\
182    & 	RX J1639.9+5347	 &    16 39 55.6  &  +53 47 56    &   0.69  $\pm$ 0.08 &  0.111 & &   &  -  & 33  \\
202    & 	RX J2137.1+0026	 &    21 37 06.7  &  +00 26 51    &   0.03  $\pm$ 0.01 &  0.051 & 0.051 & 347 $\pm$ 65  & 17  & 13  \\
209    &	RX J2239.5-0600	 &    22 39 34.4  &  -06 00 14    &   0.08  $\pm$ 0.03 &  0.173 & &   &  -  & 11  \\
217    &	RX J2319.5+1226	 &    23 19 33.9  &  +12 26 17    &   0.27  $\pm$ 0.03 &  0.126 & &   &  -  & 29  \\
      &	    &		      &		 &         &	& &	 &     &           \\
		\hline                                                       
		\hline
	\end{tabular}  
	}
	\label{table1}
\end{table*}

\begin{table*}
	\centering
	\caption{The Red Sequence linear fits. Column (1) shows the three redshift bins; columns (2) and (3), the best linear fit of the spectroscopic RS
slopes and zero-points, respectively; column (4), the  1 $\sigma$ dispersion of the spectroscopic fit; and column (5), the 1 $\sigma$ photometric dispersion. } 
 	{\small
 	\begin{tabular}{lcccc}
		\hline
		\hline
z bin &  slope    & zero-point & spec 1$\sigma$ & phot 1$\sigma$ \\
 	\hline
         &	    &		      &		 &     \\   
z$_1$  &  -0.032 $\pm$ 0.006  & 0.269 $\pm$ 0.132  & 0.148  & 0.312\\
z$_2$   & -0.035 $\pm$ 0.007  & 0.111 $\pm$ 0.166  & 0.040  & 0.093\\
z$_3$   & -0.034 $\pm$ 0.019  & 0.103  $\pm$ 0.341  & 0.100  & 0.166\\
         &	    &		      &		 &     \\
		\hline                                                       
		\hline
	\end{tabular}  
	}
	\label{table2}
\end{table*}

\begin{table*}
	\centering
	\caption{The Concentration index distributions.   Column (1) shows the three redshift bins; and  columns (2) and (3), the mean $C$ values for the two populations. }
	{\small
 	\begin{tabular}{lcc}
		\hline
		\hline
z bin &  C   & C  \\
 &  Early-types   & late-types \\

 	\hline
         &			      &		    \\   
z$_1$  &  2.95 $\pm$ 0.23   & 2.06 $\pm$ 0.23\\
z$_2$   & 2.72 $\pm$ 0.42   & 2.10 $\pm$ 0.20\\
z$_3$   & 2.57 $\pm$ 0.35   & 2.25 $\pm$ 0.33\\
         &	 		      &		     \\
		\hline                                                       
		\hline
	\end{tabular}  
	}
	\label{table3}
\end{table*}

\begin{table*}
	\centering
	\caption{The Schechter LF fits. Column (1) shows the three redshift bins; columns (2) to (4), the Schechter LF parameters using spectroscopic members; and columns (5) to (7), those with the full sample of members.}
	{\small
 	\begin{tabular}{c|c|c|c|c|c|c}
		\hline
 \hline
z bin &  & Spec members & & & All members & \\
 &  $M^*$ & $\alpha$ & $\Phi^*$ & $M^*$ & $\alpha$ & $\Phi^*$ \\
 \hline
 & & & & & & \\
z$_1$  & -21.10 $\pm$ 0.05 & -1.25 $\pm$ 0.03 & -1.49 $\pm$ 0.01 & -21.10 $\pm$ 0.07 & -1.16 $\pm$ 0.05 & -1.59 $\pm$ 0.02   \\
z$_2$  & -21.00 $\pm$ 0.09 & -0.96 $\pm$ 0.10 & -1.23 $\pm$ 0.02 & -21.00 $\pm$ 0.10 & -0.85 $\pm$ 0.04 & -1.39 $\pm$ 0.02  \\
z$_3$  & -20.71 $\pm$ 0.13 & -0.50 $\pm$ 0.11 & -0.53 $\pm$ 0.10 & -20.80 $\pm$ 0.11 & -0.76 $\pm$ 0.09 & -0.68 $\pm$ 0.12 \\
        &	 		      &		     \\
		\hline                                                    	\hline
	\end{tabular}  
	}
	\label{table4}
\end{table*}





\begin{thebibliography}{99}

\bibitem[Abazajian et al.(2009)]{abazajian2009} Abazajian, K.~N., Adelman-McCarthy, J.~K., Ag{\"u}eros, M.~A., et al.\ 2009, \apjs, 182, 543
\bibitem[Albareti et al.(2017)]{dr13} Albareti, F.~D., Allende Prieto, C., Almeida, A., et al.\ 2017, \apjs, 233, 25
\bibitem[Annunziatella et al.(2014)]{annunziatella2014} Annunziatella, M., Biviano, A., Mercurio, A., et al.\ 2014, \aap, 571, A80 
\bibitem[Baldry et al.(2004)]{baldry2004} Baldry, I.~K., Balogh, M.~L., Bower, R., Glazebrook, K., \& Nichol, R.~C.\ 2004, The New Cosmology: Conference on Strings and Cosmology, 743, 106
\bibitem[Balogh et al.(2002)]{balogh2002} Balogh, M., Bower, R.~G., Smail, I., et al.\ 2002, \mnras, 337, 256 
\bibitem[Beers et al.(1990)]{beers1990} Beers, T.~C., Preston, G.~W., Shectman, S.~A., \& Kage, J.~A.\ 1990, \aj, 100, 849
\bibitem[Biviano et al.(2013)]{biviano2013} Biviano, A., Rosati, P., Balestra, I., et al.\ 2013, \aap, 558, A1 
\bibitem[Bou{\'e} et al.(2008)]{boue2008} Bou{\'e}, G., Adami, C., Durret, F., Mamon, G.~A., \& Cayatte, V.\ 2008, \aap, 479, 335
\bibitem[Bower et al.(1992)]{bower92a} Bower, R.~G., Lucey, J.~R., \& Ellis, R.~S.\ 1992, \mnras, 254, 589 
\bibitem[Bower et al.(1992)]{bower92b} Bower, R.~G., Lucey, J.~R., \& Ellis, R.~S.\ 1992, \mnras, 254, 601
\bibitem[Bower et al.(2006)]{bower2006} Bower, R.~G., Benson, A.~J., Malbon, R., et al.\ 2006, \mnras, 370, 645
\bibitem[Christlein \& Zabludoff(2003)]{christleinzabludoff2003} Christlein, D., \& Zabludoff, A.\ 2003, \apss, 285, 197
\bibitem[Collister et al.(2007)]{collister2007} Collister, A., Lahav, O., Blake, C., et al.\ 2007, \mnras, 375, 68
\bibitem[Costa-Duarte et al.(2016)]{costaduarte2016} Costa-Duarte, M.~V., O'Mill, A.~L., Duplancic, F., Sodr{\'e}, L., \& Lambas, D.~G.\ 2016, \mnras, 459, 2539
\bibitem[de Filippis et al.(2011)]{defilippis2011} de Filippis, E., Paolillo, M., Longo, G., et al.\ 2011, \mnras, 414, 2771
\bibitem[Dekel \& Silk(1986)]{dekelsilk1986} Dekel, A., \& Silk, J.\ 1986, \apj, 303, 39
\bibitem[de Lapparent(2003)]{delapparent2003} de Lapparent, V.\ 2003, \aap, 408, 845
\bibitem[De Lucia et al.(2004)]{delucia2004} De Lucia, G., Poggianti, B.~M., Arag{\'o}n-Salamanca, A., et al.\ 2004, \apjl, 610, L77 
\bibitem[De Lucia et al.(2010)]{delucia2010} De Lucia, G., Boylan-Kolchin, M., Benson, A.~J., Fontanot, F., \& Monaco, P.\ 2010, \mnras, 406, 1533 
\bibitem[Doi et al.(2010)]{doi2010} Doi, M., Tanaka, M., Fukugita, M., et al.\ 2010, \aj, 139, 1628
\bibitem[Gaidos(1997)]{gaidos1997} Gaidos, E.~J.\ 1997, \aj, 113, 117
\bibitem[Garilli et al.(1999)]{garilli1999} Garilli, B., Maccagni, D., \& Andreon, S.\ 1999, \aap, 342, 408 
\bibitem[Gilbank \& Balogh(2008)]{gil08} Gilbank, D.~G., \& Balogh, M.~L.\ 2008, \mnras, 385, L116
\bibitem[Gladders et al.(1998)]{gladders1998} Gladders, M.~D., L{\'o}pez-Cruz, O., Yee, H.~K.~C., \& Kodama, T.\ 1998, \apj, 501, 571
\bibitem[Gladders \& Yee(2005)]{gladders2005} Gladders, M.~D., \& Yee, H.~K.~C.\ 2005, \apjs, 157, 1 
\bibitem[Gonzalez et al.(2015)]{gonzalez2015} Gonzalez, E.~J., Fo{\"e}x, G., Nilo Castell{\'o}n, J.~L., et al.\ 2015, \mnras, 452, 2225 
\bibitem[Goto et al.(2002)]{goto2002} Goto, T., Okamura, S., McKay, T.~A., et al.\ 2002, \pasj, 54, 515
\bibitem[Jenkins et al.(2001)]{jenkins2001} Jenkins, A., Frenk, C.~S., White, S.~D.~M., et al.\ 2001, \mnras, 321, 372
\bibitem[Kauffmann \& Charlot(1998)]{kauffmanncharlot1998} Kauffmann, G., \& Charlot, S.\ 1998, \mnras, 294, 705
\bibitem[Kauffmann et al.(2003a)]{kauff3a} Kauffmann, G., Heckman, T.~M., White, S.~D.~M., et al.\ 2003a, \mnras, 341, 33
\bibitem[Kauffmann et al.(2003b)]{kauff3b} Kauffmann, G., Heckman, T.~M., Tremonti, C., et al.\ 2003b, \mnras, 346, 1055
\bibitem[Klypin et al.(1999)]{klypin1999} Klypin, A., Gottl{\"o}ber, S., Kravtsov, A.~V., \& Khokhlov, A.~M.\ 1999, \apj, 516, 530
\bibitem[Lan et al.(2016)]{lan2016} Lan, T.-W., M{\'e}nard, B., \& Mo, H.\ 2016, \mnras, 459, 3998
\bibitem[Lerchster et al.(2011)]{lerchster2011} Lerchster, M., Seitz, S., Brimioulle, F., et al.\ 2011, \mnras, 411, 2667
\bibitem[Lopez-Cruz(1997)]{lopezcruz1997} Lopez-Cruz, O.\ 1997, Ph.D.~Thesis, 2803
\bibitem[McGee et al.(2009)]{mcgee2009} McGee, S.~L., Balogh, M.~L., Bower, R.~G., Font, A.~S., \& McCarthy, I.~G.\ 2009, \mnras, 400, 937
\bibitem[Moore \& Ridge(1999)]{moore1999} Moore, T., \& Ridge, N.\ 1999, Star Formation 1999, 293
\bibitem[Mullis et al.(2003)]{mull03} Mullis, C.~R., McNamara, B.~R., Quintana, H., et al.\ 2003, \apj, 594, 154 
\bibitem[Nilo Castell{\'o}n et al.(2014)]{nilo14} Nilo Castell{\'o}n, J.~L., Alonso, M.~V., Lambas, D.~G., et al.\ 2014, \mnras, 437, 2607
\bibitem[Nilo Castell{\'o}n et al.(2016)]{nilo16} Nilo Castell{\'o}n, J.~L., Alonso, M.~V., Garc{\'{\i}}a Lambas, D., et al.\ 2016, \aj, 151, 151
\bibitem[O'Mill et al.(2011b)]{omill11} {O'Mill}, A.~L., {Duplancic}, F. and {Garc{\'{\i}}a Lambas}, 2011, \mnras, 413, 1395
\bibitem[Paolillo et al.(2000)]{paolillo2000} Paolillo, M., Andreon, S., Longo, G., et al.\ 2000, \memsai, 71, 1069
\bibitem[Paolillo et al.(2001)]{paolillo2001} Paolillo, M., Andreon, S., Longo, G., et al.\ 2001, \aap, 367, 59
\bibitem[Piffaretti et al.(2011)]{piffaretti2011} Piffaretti, R., Arnaud, M., Pratt, G.~W., Pointecouteau, E., \& Melin, J.-B.\ 2011, \aap, 534, A109
\bibitem[Popesso et al.(2006)]{popesso2006} Popesso, P., Biviano, A., B{\"o}hringer, H., \& Romaniello, M.\ 2006, \aap, 445, 29
\bibitem[Sarron et al.(2018)]{sarron2018} Sarron, F., Martinet, N., Durret, F., \& Adami, C.\ 2018, \aap, 613, A67 
\bibitem[Schechter(1976)]{schecter1976} Schechter, P.\ 1976, \apj, 203, 297
\bibitem[Schmidt(1968)]{vmax} Schmidt, M.\ 1968, \apj, 151, 393 
\bibitem[Stott et al.(2009)]{stott2009} Stott, J.~P., Pimbblet, K.~A., Edge, A.~C., Smith, G.~P., \& Wardlow, J.~L.\ 2009, \mnras, 394, 2098
\bibitem[Stoughton et al.(2002)]{stout} Stoughton, C., Lupton, R.~H., Bernardi, M., et al.\ 2002, \aj, 123, 485
\bibitem[Strateva et al.(2001)]{strateva2001} Strateva, I., Ivezi{\'c}, {\v Z}., Knapp, G.~R., et al.\ 2001, \aj, 122, 1861 
\bibitem[Valotto et al.(1997)]{valotto1997} Valotto, C.~A., Nicotra, M.~A., Muriel, H., \& Lambas, D.~G.\ 1997, \apj, 479, 90
\bibitem[Vikhlinin et al.(1998)]{vikhlinin1998} Vikhlinin, A., McNamara, B.~R., Forman, W., et al.\ 1998, \apjl, 498, L21
\bibitem[Visvanathan \& Sandage(1977)]{vis77} Visvanathan, N., \& Sandage, A.\ 1977, \apj, 216, 214
\bibitem[Zandivarez \& Mart{\'{\i}}nez(2011)]{zandivarez2011} Zandivarez, A., \& Mart{\'{\i}}nez, H.~J.\ 2011, \mnras, 415, 2553 


\end{thebibliography}








\bsp	
\label{lastpage}
\end{document}